\documentclass[aps,showpacs,prc,nofootinbib,floatfix,]{revtex4}

\usepackage{graphicx}

\usepackage{latexsym}
\usepackage{hhline}
\usepackage{amssymb}
\usepackage{wasysym}
\usepackage{epsfig}
\usepackage{ulem}

\sloppypar

\newcommand{\be}{\begin{equation}}
\newcommand{\ee}{\end{equation}}
\newcommand{\bea}{\begin{eqnarray}}
\newcommand{\eea}{\end{eqnarray}}
\newcommand{\ba}{\begin{array}}
\newcommand{\ea}{\end{array}}
\newcommand{\bi}{\begin{itemize}}
\newcommand{\ei}{\end{itemize}}

\newcommand{\foh}{\frac{1}{2}}

\begin{document}

\title{ $\eta$ photoproduction in the resonance energy region.
\footnote{Supported by Forschungszentrum Juelich}}

\author{V. Shklyar
}
\email{shklyar@theo.physik.uni-giessen.de}
\author{H. Lenske}
\author{U. Mosel}
\affiliation{Institut f\"ur Theoretische Physik, Universit\"at Giessen, D-35392
Giessen, Germany}


\begin{abstract}
The $\eta$ production in the nucleon resonance energy region is studied within the unitary 
coupled-channels effective Lagrangian approach of the Giessen model. 
We demonstrate that the second peak recently  observed 
in  the cross section of  $\eta$ photoproduction on the neutron at $\sqrt{s}$=1.66\,GeV can be 
explained in terms of
coupled-channel effects due to $S_{11}(1650)$ and $P_{11}(1710)$ resonance excitations.
\end{abstract}


\maketitle
Most of the information about the electromagnetic properties
of nucleon resonances comes from the analysis of  pion photoproduction data.
However, since we have to expect that not all of the resonances couple equally strong to 
the $\pi N$ channel  other production and decay scenarios must be investigated.
Such a  complementary information on  resonance spectra can be obtained from the 
study  of   reactions with $K\Lambda$, $\eta N$, $\omega N$ etc. in the 
final state. Recently, $K\Lambda$ photoproduction has attracted  considerable  attention 
\cite{Mart:1999,Mart:2006dk,Julia-Diaz:2006is,shklyar:2005c} in prospects of searching for
'hidden' states predicted in 
\cite{Capstick:1998uh,Capstick:1993kb,Capstick:1992uc}.
The $\eta$ production on the  neutron might be of particular interest for the search
of narrow 'hidden' states  predicted by  some quark models 
\cite{Diakonov:1997mm,Jaffe:2003sg}.
Due to the isoscalar nature of the  $\eta$  meson this reaction selects only 
isospin-$\foh$ channel which also simplifies the analysis.
The previous experimental studies 
of $\gamma d$ scattering \cite{Krusche:1995,Weiss:2002} 	
have shown that  $\eta$ photoproduction on the neutron at c.m. energies up to {\it $\sqrt{s}$=1.6}\,GeV is 
governed by the excitation of the $S_{11}(1535)$ resonance.
Recently, the GRAAL \cite{Kuznetsov:2006kt} and CBELSA-TAPS  \cite{Krusche:2006}
collaborations reported on their preliminary $\gamma n\to\eta n$ data
which has been extracted from the analysis of $\gamma d$ scattering. 
Although these measurements need to be confirmed
a striking and unexpected result of these findings is that the integrated neutron cross  section 
has an additional maximum at c.m. energy  around {\it  $\sqrt{s}$=1.66 GeV}. 
To our knowledge a 
consistent explanation of this phenomenon is pending.

The central question is whether the observed structure comes from the excitation of an unknown 
baryonic state. 
In  \cite{Arndt:2003ga}  the existence of a narrow resonance with the mass  $M$=1.675\,GeV
and strong coupling to the $\eta n$ final state has been predicted.
The contribution from this resonance has been
proposed \cite{Azimov:2005jj} 
as an explanation of the peak seen in  the preliminary $\gamma n^*\to \eta n$  
GRAAL  data \cite{Kuznetsov:2006kt}  at $\sqrt{s}=$1.6...1.7 GeV. However, before any 
conclusion can be drawn the conventional mechanisms in the $\eta$ photoproduction 
both on the proton and on the neutron has to  be investigated in detail.
In this letter we present a first attempt for a multichannel analysis
of $\eta p$ and $\eta n$ reactions within the Giessen Model, a unitary 
coupled-channel  approach taking into account constraints  from the other 
scattering channels.
The Giessen model \cite{shklyar:2004a,Penner:2002a,Penner:2002b} has been developed 
for the simultaneous analysis of the pion- and 
photon-induced reactions up to about \mbox{2 GeV}. 
In \cite{shklyar:2004b,shklyar:2005c} an updated solution of the coupled-channel 
problem has been obtained to 
$\pi N \to \pi N$,\,$2\pi N$,\,$\eta N$,\,$\omega N$,\,$K \Lambda$,\,$K\Sigma$ and 
$\gamma N \to \gamma N$,\,$\pi N$,\,$\eta N$,\,$\omega N$,\,$K \Lambda$,\,$K\Sigma$ 
reactions at energies from the threshold up to 2\,GeV. 
At that time  the $\eta$ photoproduction on the neutron
was beyond the scope of the calculations. 
Since the hadronic resonance  parameters
have been extracted in \cite{shklyar:2004b,shklyar:2005c} the  extension of the model
to $\gamma n \to \eta n$ is in principle straightforward if the neutron helicity amplitudes of all 
resonances   would be known.  
The electromagnetic properties of most of the discovered states
were extracted only from the analysis of the pion photoproduction data. Hence, the 
uncertainties of the extracted parameters would bring  a large  ambiguity to such a 
calculations.

A very important example relevant for the case at hand  
are the electromagnetic properties of the $S_{11}(1535)$  resonance.  Because of the strong coupling 
of  this state to $\eta N$, the amplitude ratio $R= A^n_{1/2}(S_{11}(1535))/A^p_{1/2}(S_{11}(1535))$   
defines a balance between $\eta$ meson  photoproduction on the neutron and the proton  at 
energies close to this resonance mass.
The various analyses of pion photoproduction find  this ratio  being spread  over a wide range 
$R=-0.3$\,...$-1$ (  see  the Particle Data Group (PDG) \cite{PDBook} and references therein).
On the other hand the  combined studies
of the $\eta p$  and $\eta n$  photoproduction data 
\cite{Sauermann:1997,Drechsel:1998,Mukhopadhyay:1995,Chiang:2002vq}
seems to agree on the value \mbox{$R\approx-0.8$}.
Hence, one can hope that the  multichannel analysis of the $\pi N$- and $\eta N$-channels
maximally constrains the resonance  parameters, thereby  a solid conclusion on the reaction 
mechanism  and resonance parameters can be drawn. 

The aim of this letter is as follows. We repeat our previous calculations 
\cite{shklyar:2004b,shklyar:2005c} but taking into account
presently available $\eta n$ data to constrain 
the couplings  $N^*\to \gamma n$.
Using the hadronic parameters from \cite{shklyar:2004b} we predict the 
angular distribution
and beam asymmetry for \mbox{$\gamma n \to \eta n $} scattering in the energy region 
where the second peak in the $\eta n$ data is observed.

\begin{figure}
{\includegraphics*[width=14cm]{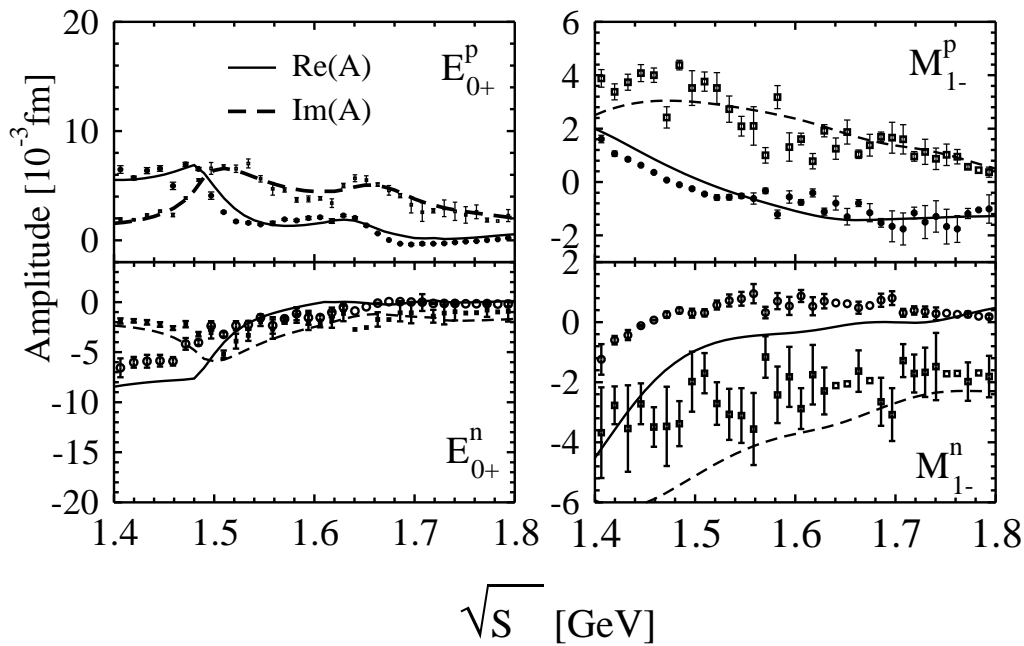}}
       \caption{  Multipoles for pion photoproduction: proton (top) and neutron (bottom) respectively.
   Single energy amplitudes and energy-dependent solutions of the GW 
analysis \cite{Arndt:1996} are shown by circles (real) and 
   squares (imaginary) respectively.
      \label{multipoles}}
\end{figure}

In \cite{shklyar:2004b}  $\eta n$ data were not included in the fit. 
In the present calculation we include the experimental data on the ratio 
$(d\sigma/d\Omega)_n/(d\sigma/d\Omega)_p$   of neutron to proton $\eta$ photoproduction 
cross sections from \cite{Hoffmann-Rothe:1997sv}. 
These data cover the  energy region  1.5...1.6 GeV but with  large statistical errors.
We include these data in our fit but multiply the original error bars  
by  the factor of $\frac{1}{3}$.  
Above 1.6 GeV we include the preliminary data points of the total 
cross section $\gamma n \to \eta n$ from \cite{jaegle:prelim}. 
Starting from our
best solution to the pion- and photon-induced reactions we perform an additional  fit varying
only the helicity decay amplitudes  and $\eta NN^*$-couplings of the isospin-$\foh$ 
resonances  keeping all other parameters fixed. 
The obtained parameters are shown in \mbox{Table \ref{param}} in comparison with the results
from \cite{shklyar:2004b}.
The variation of the $\eta NN^*$-couplings
leads in general  to a modification of total  widths of the resonances  and affects 
the description of $\gamma p\to (\pi/\eta) p$ 
reaction. Hence a variation  of the $A^{p}_{1/2}$ amplitudes was also allowed. 
However, only tiny  changes in the proton helicity amplitudes are observed, indicating the 
reliability of the former calculations and their stability.
The corresponding $S_{11}$- and $P_{11}$-multipoles for  pion-photoproduction 
shown in  Fig.~\ref{multipoles}  in comparison with the energy-independent solution 
from the analysis of the group at the  George-Washington (GW) university \cite{Arndt:1996}. 
At those energies, where the 
energy-independent solutions are not available, we have used the 
energy-dependent solutions from \cite{Arndt:1996}.

\begin{table}[t]
 \begin{tabular}{clllllllll}
  \hline
 \hline
  $N^*$ & mass & $\Gamma_{tot}$ &$\Gamma_{\pi N}$ & $\Gamma_{\eta N}$ & $A^{p}_{1/2}$  &
   $A^{n}_{1/2}$ & $A^{p}_{\frac{3}{2}}$ &
 $A^{n}_{\frac{3}{2}}$ \\
 \hline
 $S_{11}$(1535)
& 1526 & 136 & 34.4 & $56.2[+]$            &  95 &  -74 & \multicolumn{2}{c}{---} \\
& 1526 & 136 & 34.4 & $56.1[+]$            &  92 &  -13 & \multicolumn{2}{c}{---}  \\ 
& 1535(10) &  150(25) &  45(10)   & 53(7)  &  90(30)   &  -46(27)   & \multicolumn{2}{c}{---}  \\ 
%
 $S_{11}$(1650)
& 1664     & 133      & 71.9    & $ 2.5[-]$ &   57     &   -9        & \multicolumn{2}{c}{---} \\
& 1664     & 131      & 72.4    & $ 1.4[-]$ &   57     &  -25        & \multicolumn{2}{c}{---}\\ 
& 1655(15) & 165(20)  & 72(22 ) & 6(3)      &   53(16) &  -15(21)    & \multicolumn{2}{c}{---}  \\ 
 \hline
 $P_{11}$(1440)
& 1517       & 608      & 56.0       & --- &  -84    &  138     & \multicolumn{2}{c}{---}  \\
& 1517       & 608      & 56.0       & --- &  -84    &  138     & \multicolumn{2}{c}{---} \\ 
& 1440(30)   & 325(125) & 65(10)     & --- &  -65(4) &  40(10)  & \multicolumn{2}{c}{---}  \\ 
 $P_{11}$(1710)
& 1723       & 397       &  1.7      & $41.5[+]$ &  -50    &   24   & \multicolumn{2}{c}{---} \\
& 1723       & 408       &  1.7      & $43.0[+]$ &  -50    &   68   & \multicolumn{2}{c}{---} \\ 
& 1710(10)   &  150(100) &  15(5)    &  6(1)     &   9(22) & -2(14) & \multicolumn{2}{c}{---}  \\ 
 \hline
 $P_{13}$(1720)
& 1700       & 152       & 17.1      & $ 0.1$[+] &  -65    &    3    &   35     &   -1    \\
& 1700       & 152       & 17.1      & $ 0.2$[+] &  -65    &    1    &   35     &   -4     \\ 
& 1725(25)   & 225(75)   & 15(5)     & 4(1)      &   18(30)&    1(15)&  -19(20) &  -29(61)  \\ 
 $P_{13}$(1900)
& 1998       & 369       & 24.5      & $ 5.4$[-] &   -8    &   12    &    0     &   23 \\
& 1998       & 404       & 22.2      & $ 2.5$[-] &   -8    &  -19    &    0     &    6 \\
& 1900       & NG        &  26(6)    &  14(5)    &  -17    &  -16    &  31      &    2  \\ 
 \hline
 $D_{13}$(1520)
& 1505       & 100       & 56.5      & $ 1.2$[+]   &  -15    &  -64    &  146     & -136 \\
& 1505       & 100       & 56.6      & $ 1.2$[+]   &  -13    &  -70    &  145     & -141 \\ 
& 1520(5)    & 112(12)   &  60(5)    &   0.2(0.04) &  -24(9) &  -59(9) &  166(5)  & -139(11)\\

 $D_{13}$(1950)
& 1934       & 855       & 10.5      & $ 0.1$[-]   &   11    &   26    &   26     &  -55 \\
& 1934       & 859       & 10.5      & $ 0.5$[-]   &   11    &   40    &   26     &  -33 \\
& 2080       & NG        & NG        & 4(4)        & -20(8)  &   7(13) & 17(11)   &  -53(34)\\
 \hline
 $D_{15}$(1675)
& 1666       & 148       & 41.1      & $ 0.1$[+]   &    9    &  -56    &   21     &  -84 \\
& 1666       & 148       & 41.1      & $ 0.3$[+]   &    9    &  -56    &   21     &  -84 \\ 
& 1675(5)    & 146(16)   & 40(5)     & 0(1)        & 19(8)   &  -43(12)& 15(9)    &  -58(13)\\
 \hline
 $F_{15}$(1680)
& 1676       & 115       & 68.3      & $ 0.0$[+]   &    3    &   30    &  116     &  -48 \\
& 1676       & 115       & 68.3      & $ 0.0$[+]   &    3    &   30    &  116     &  -48  \\ 
& 1685(5)    & 130(10)   & 68(3)     &  0(1)       & -15(6)  &   29(10)&  133(12) &  -33(9) \\
 $F_{15}$(2000)
& 1946       & 198       &  9.9      & $ 2.0$[-]   &   11    &    9    &   25     &   -3 \\
& 1946       & 198       &  9.9      & $ 2.0$[-]   &   11    &    9    &   25     &   -3 \\ 
& 2000       & 490(130)  &  8(5)     &  NG         &         &         &          &         \\
 \hline 
 \hline
   \end{tabular}
  \caption{
Parameters of resonances considered in the present work.
 First line: parameters obtained in the present calculations.
 Second line: parameters are taken from our previous analysis \cite{shklyar:2004b}.
 In square brackets the sign of the $\eta NN^*$ coupling relative to the
 $\pi NN^*$ coupling is given.
 Third line: values from PDG; in brackets estimated errors are given. 
 NG - no average value in PDG is given.
 Helicity amplitudes  are given in units of  10$^{-3}$GeV$^{-\foh}$.
    \label{param}} 
\end{table}

Including the new $\eta n$ data we find the most significant change 
in the neutron helicity amplitude for
the  $N_{1535}^*$ resonance.
The present value $A^n_{1/2}(S_{11}(1535))=-74\times 10^{-3}$GeV$^{-1/2}$ 
is much larger in magnitude then the one determined previously \cite{shklyar:2004b}.
The change in the decay amplitude is necessary to  describe the data \cite{Hoffmann-Rothe:1997sv} 
in the energy region 1.5...1.6 GeV, where the $S_{11}(1535)$ state gives the major contributions
to $\eta$ photoproduction. 
It leads to the  ratio  
\mbox{$R= A^n_{1/2}(S_{11}(1535))/A^p_{1/2}(S_{11}(1535))\approx-0.8$} which is in line with the
conclusions obtained in \cite{Sauermann:1997,Drechsel:1998,Mukhopadhyay:1995,Chiang:2002vq}.
At the  same time, the changes in the properties of $S_{11}(1535)$ affect  the real part
of the $E_{0+}^n$ multipole around 1.53 GeV, see Fig.~\ref{multipoles}. Interestingly, the same 
behaviour has also been found in the  MAID   calculations \cite{Drechsel:1998}.

\begin{figure}
{\includegraphics*[width=7cm]{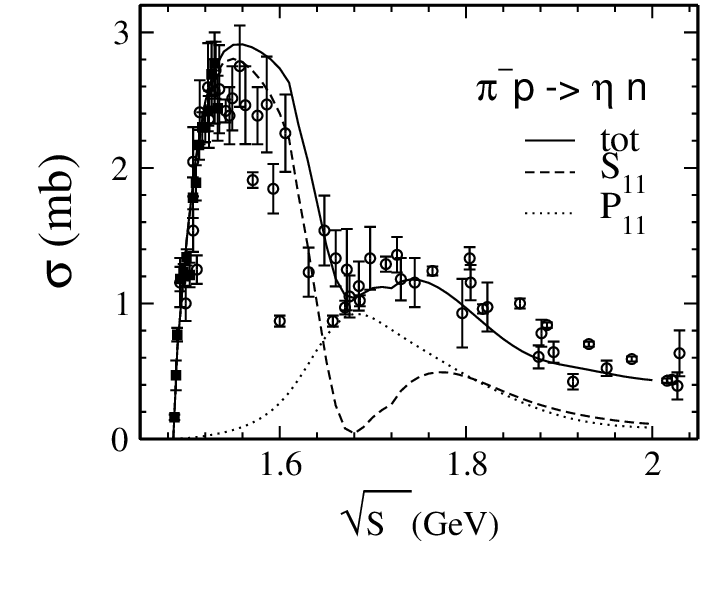}}
       \caption{ $\pi^- p \to \eta n$ total and partial wave cross sections.
      The experimental data are taken from\,\cite{Penner:PhD}. 
      \label{piN_etaN}}
\end{figure}

\begin{figure}
{\includegraphics*[width=7cm]{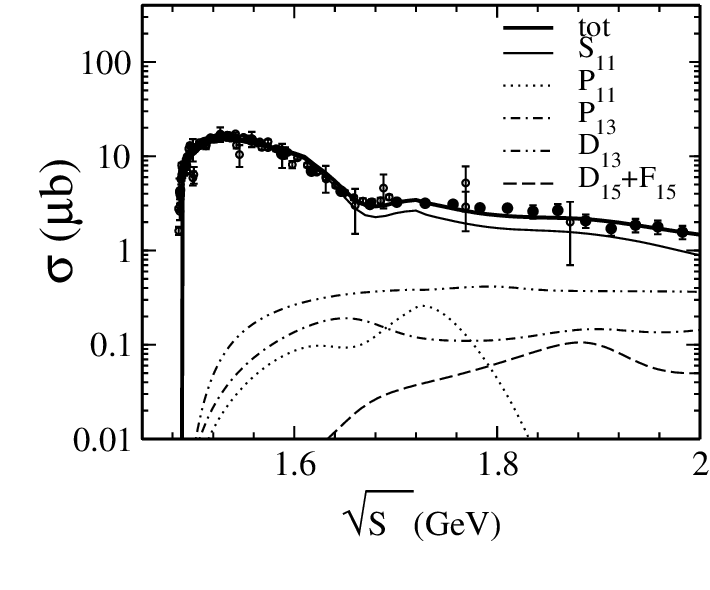}}
       \caption{ $\gamma p \to \eta p$ total and partial wave cross sections.
The experimental data are taken from \cite{CBELSA:2005,CLAS:2002,GRAAL:2002}.
      \label{eta-p}}
\end{figure}

 \begin{figure}
  \includegraphics*[width=7cm]{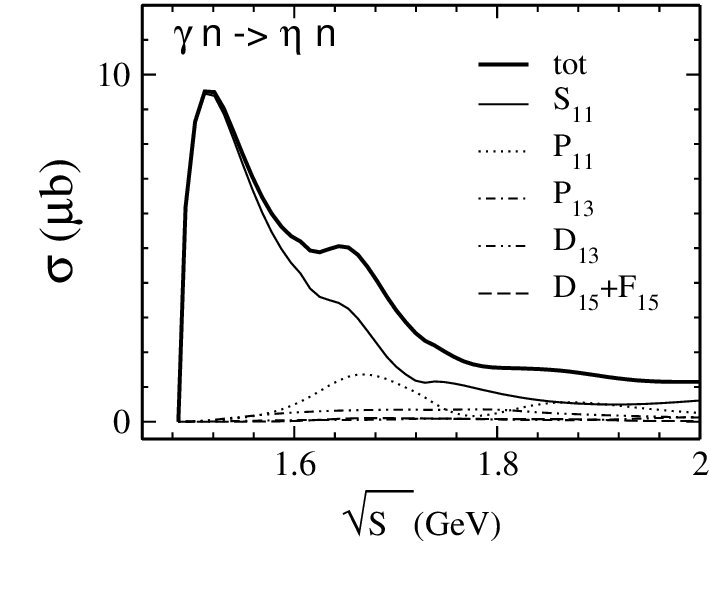}
       \caption{  $\gamma n \to \eta n$  total and partial wave cross sections. The kinks at
 {1.61} GeV and \mbox{1.72 GeV}  are the threshold effects coming from $K\Lambda$ and $\omega N$.     
      \label{eta-n1}}
\end{figure}

 \begin{figure}
\begin{minipage}{8cm}
 {\includegraphics*[width=7.5 cm]{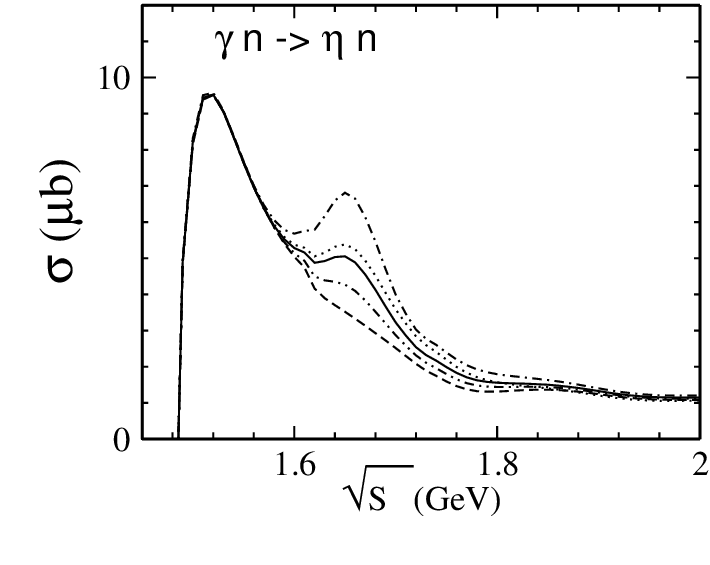}}
\end{minipage}
\begin{minipage}{8cm}
 {\includegraphics*[width=7.5 cm]{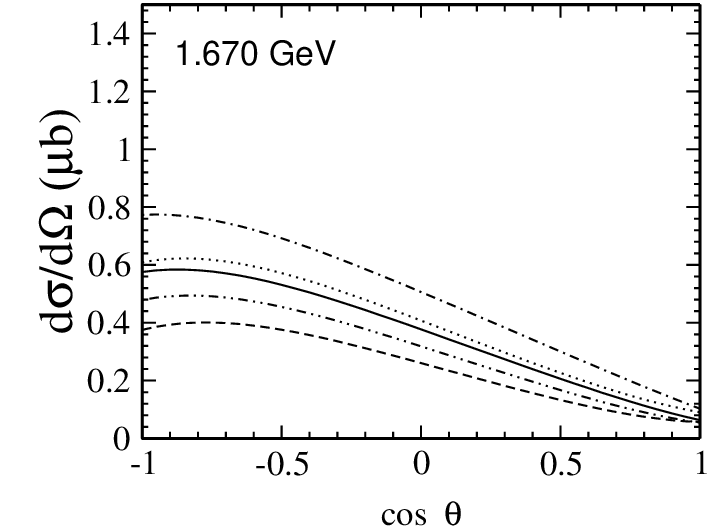}}
 \end{minipage}
       \caption{  $\gamma n \to\eta n$ total~(left) and differential~(right) cross 
sections calculated using the parameter set from 
 Table \ref{param} (solid line) and with  different choice of the
 neutron helicity amplitudes for the $S_{11}(1650)$ and $P_{11}(1710)$ resonances:
 $A^n_{1/2}(S_{11}(1650))$=-24 (dashed),   $A^n_{1/2}(S_{11}(1650))$=-16 (dashed-double-dotted),
 $A^n_{1/2}(S_{11}(1650))$=+3 (dashed-dotted), $A^n_{1/2}(P_{11}(1710))$=+17 (dotted), where
 the helicity amlitudes are given in units of $10^{-3}$GeV$^{-\foh}$.
      \label{eta-n2}}
 \end{figure}

The second $S_{11}(1650)$ state has a large 
branching ratio to $\pi N$ thereby a clear resonance behaviour is seen
in the proton electric multipole  $E_{0+}^p$ at the energy 1.66 GeV.
The effect from this 
resonance is much less pronounced in the $E_{0+}^n$ multipole which points to
a small magnitude of the corresponding neutron  helicity amplitude. 
Note, that in  the literature there is even no agreement on the sign of $A^n_{1/2}(S_{11}(1650))$,
PDG \,\cite{PDBook}. 
We obtain  
$A^n_{1/2}(S_{11}(1650))=-0.009$~GeV$^{-\foh}$ {which is close             
to the} value found by Arndt et. al. \cite{Arndt:1996} and Penner and Mosel \cite{Penner:2002b}.
For the present discussion also the  $P_{11}(1710)$ resonance is important but the 
properties of this state are  not well  determined, \cite{PDBook}. 
In our previous study \cite{shklyar:2004b} this resonance was found   
to have a large branching ratio to $\eta N$.
{In the present calculations the three states $S_{11}(1535)$, $S_{11}(1650)$, 
and $P_{11}(1710)$ give the major 
contributions to the $\pi^- p \to \eta n$ reaction, see  Fig.~\ref{piN_etaN}.
Above \mbox{1.65 GeV} the destructive interference with the second $S_{11}(1650)$ resonance               
decreases the effective contribution to the $S_{11}$ partial wave. A                  
similar behaviour has also been found in the calculations of
the rather different approach of the  J\"ulich                
group \cite{Gasparyan:2003fp}. The  $P_{11}(1710)$ resonance together with the background 
contributions dominate the $\pi^-p \to \eta n$ reaction in the energy region under discussion 
developing a peak in the $P_{11}$-wave cross section around 1.7 GeV.

In the present calculations the $\eta$ photoproduction on the proton is almost solely 
determined by the 
contribution from the $S_{11}$ partial wave, see Fig.~\ref{eta-p}. Other partial wave 
cross sections have very small magnitudes, they are  shown in the same figure  for completeness. 
The situation changes for the $\eta$ photoproduction on the neutron, Fig.~\ref{eta-n1}. 
In the energy region 1.5...1.6 GeV the $S_{11}(1535)$ state strongly influences
the production cross section.
The second peak in the $S_{11}$-partial wave cross section 
around 1.66 GeV stems from  the  $S_{11}(1650)$ resonance.
 We also find an additional contribution coming from 
$P_{11}(1710)$.  
In our earlier photoproduction studies \cite{Feuster:1998b,Penner:2002b} it has been concluded that
strong interference effects must exist in this kind of reactions. Hence, the resonance contributions
to the photoproduction depend strongly both on the sign and the magnitude of the $N^*$ decay parameters.
The calculated contribution from $P_{11}(1710)$ to the $\eta$ photoproduction on the proton is 
small due to destructive interference \cite{Penner:2002b}. Changing the sign of the  
$A^n_{1/2}(P_{11}(1710))$ amplitude enhances the effect of this resonance to the $\gamma n \to \eta n$ reaction. 

The two kink structures seen in the $S_{11}$ partial wave cross section 
at \mbox{1.61} GeV and \mbox{1.72 GeV} in Fig.~\ref{eta-n1} are the threshold effects coming from     
the opening of the $K\Lambda$ and $\omega N$ channels,  respectively. The effect from the $\omega N$ threshold 
is also seen in the $\gamma p \to \eta p$ cross section at 1.72 GeV, see  Fig.~\ref{eta-p}.
 Note, that taking the final width of the $\omega$ meson into account would smear out this effect.

 \begin{figure}
 {\includegraphics*[width=7.5 cm]{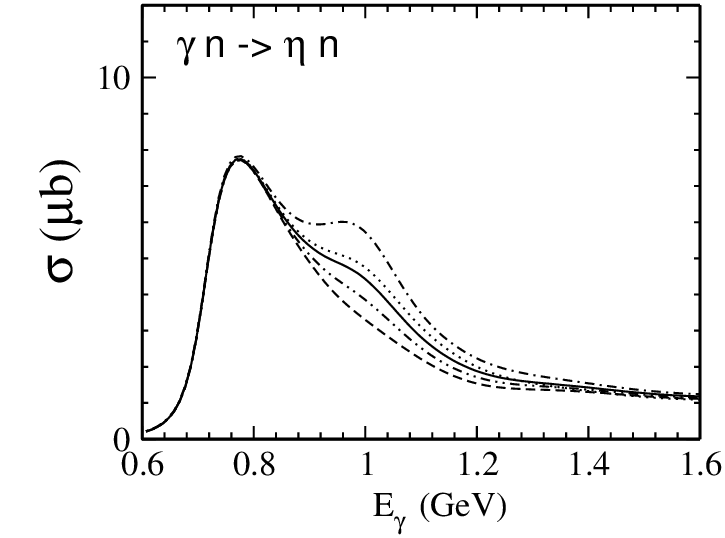}}
       \caption{ The cross sections as in the left part of Fig.~\ref{eta-n2} but smeared out over the Fermi
motion inside the deuteron. 
Notation is same as in Fig.~\ref{eta-n2}. 
      \label{etan_fermi}}
 \end{figure} 
\begin{figure}
{\includegraphics*[width=7cm]{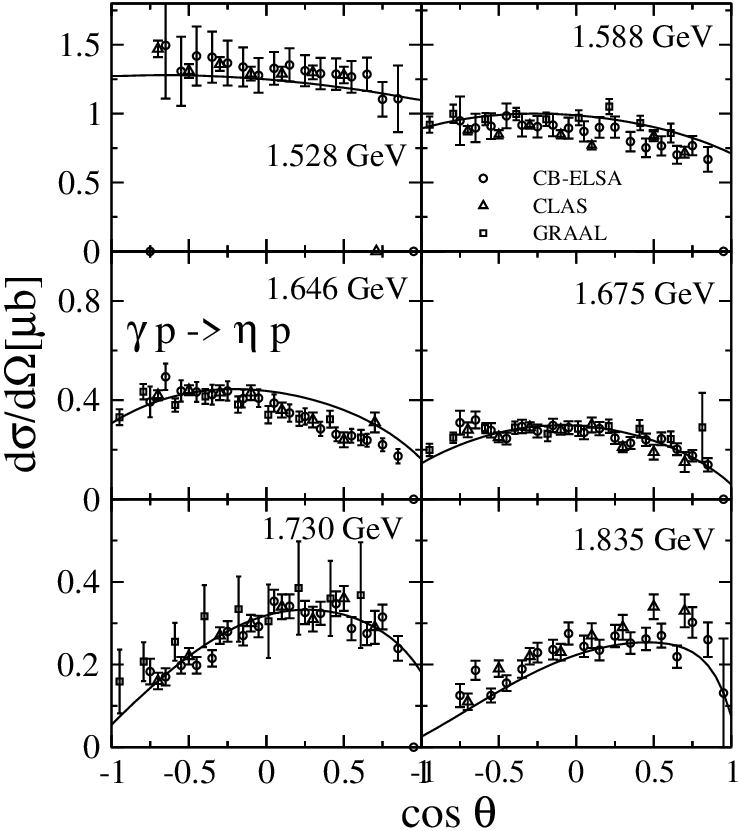}}
{\includegraphics*[width=7cm]{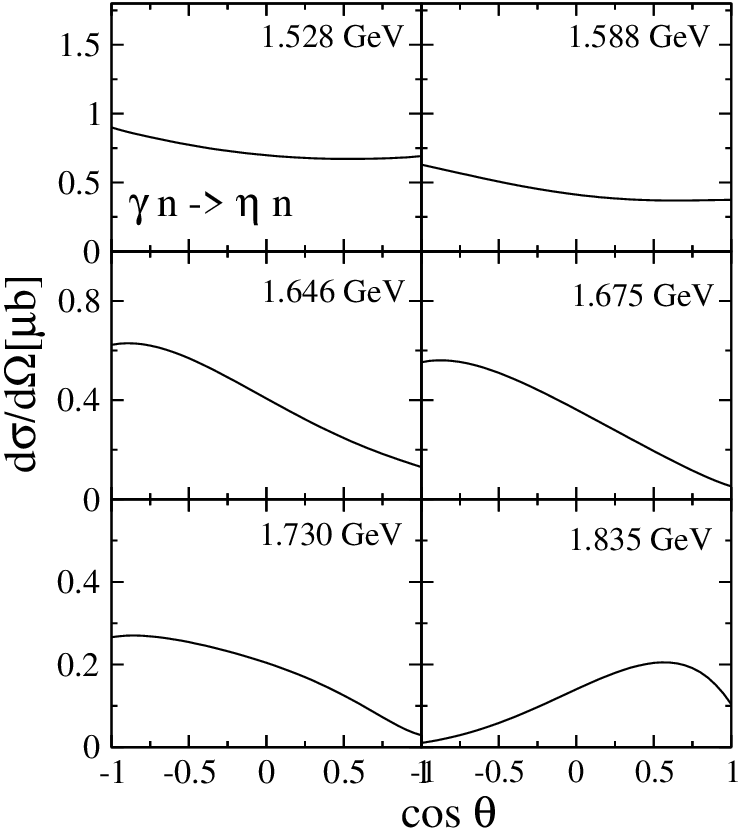}} 
       \caption{ $\gamma p \to\eta p$~(left) and  $\gamma n \to\eta n$~(right) 
differential cross sections calculated for different total c.m. energies. 
The experimental data are taken from \cite{CBELSA:2005}(CB-ELSA),
\cite{CLAS:2002}(CLAS),  and \cite{GRAAL:2002}(GRAAL).
      \label{eta-diff}}
\end{figure}

The overall magnitude of the second peak in the $\gamma n\to \eta n$ total cross section
is very sensitive to the value of the neutron  helicity amplitude of the $S_{11}(1650)$ and 
$P_{11}(1710)$ states. In the present calculations  these parameters are constrained by the preliminary 
$\gamma n^* \to \eta n$  quasi-free total cross section data  \cite{jaegle:prelim} which, however, 
has large statistical errors. These uncertainties might also affect the resonance helicity
amplitudes. Therefore,
in Fig.~\ref{eta-n2} we show  the dependence of the cross section on the
choice of parameters. 
Choosing $A^{n}_{1/2}(S_{11}(1650))=-0.024$GeV$^{-\foh}$ 
changes the interference pattern in the $S_{11}$-partial wave and strongly decreases the 
magnitude of the second peak, see the dashed curve in Fig.~\ref{eta-n2}. 
The contribution from the $S_{11}(1650)$ resonances becomes more pronounced when choosing a
different sign
for $A^{n}_{1/2}(S_{11}(1650))$. The same tendency is observed for the  $P_{11}(1710)$ resonance:
decreasing  $A^{n}_{1/2}(P_{11}(1710))$ in magnitude  enhances the overall contributions
to the integrated cross section at 1.66 GeV. The differential cross sections 
calculated at 1.67 GeV using  a different choice of the resonance parameters 
are shown in the right panel of 
Fig.~\ref{eta-n2}. It is remarkable that in all cases mainly the magnitude of the 
backward angular distribution is affected but leaving the shape almost unchanged.

The experimental information on the $\gamma n \to \eta n $  reaction is  commonly 
extracted from photo-nuclear reactions on the deuteron 
{where the proton is taken to be a spectator.} 
To take the effect of Fermi motion into account we have used the  deuteron  wave functions
obtained with the  Paris $NN$-potential \cite{Lacombe:1981eg}. Assuming
$\sigma_{\eta\, n}^\mathrm{onshell}(s) \approx \sigma_{\eta\,n}^\mathrm{offshell}(s)$ 
the effect of the Fermi smearing 
can be approximately calculated 
by folding the elementary free space eta-n cross section by the deuteron momentum distribution
\bea
\sigma(E_\gamma)=\frac{1}{4\pi}\int d^3p \biggl( |u_s(p)|^2+|u_d(p)|^2 \biggl) 
\sigma_{\eta\, n}(s^*), \,\,\,\, s^* = (k_\gamma + p)^2,
\eea
 where the information on the Fermi motion is carried by s- and d-state components
 $u_s(p)$ and $u_d(p)$ of the deutron wave functions respectively. 

The result is shown in   Fig.~\ref{etan_fermi}. 
Once the Fermi motion 
is included the overall magnitude of the  second  resonance peak becomes less pronounced but is still 
visible. A narrow resonance proposed first in \cite{Arndt:2003ga,Azimov:2005jj} 
(and  discussed more recently in 
\cite{Fix:2007st}) or the contribution from the $D_{15}(1675)$ resonance proposed by Tiator 
\cite{Baru:2006hy} have been suggested as  alternative explanations for the second peak  found 
in the  experimental data  on the $\eta n$ photoproduction \cite{Kuznetsov:2006kt,Krusche:2006}.
From our calculations we conclude instead that coupled-channel and interference effects are 
a possible explanation for the structure seen in the experimental data. However, a fair statement 
is that at present the persisting experimental uncertainties translate into corresponding 
uncertainties in the theoretically derived parameters. 
In view of these problems a final answer must wait  until more precise data are available.
It is interesting to note that the  experimental  setup  of the TAPS detector \cite{Weiss:2002} 
 apparently might  minimize or even eliminate 
the effect of Fermi smearing from the data analysis. If this indeed could be achieved
the observables shown in Fig.\ref{eta-n2} could be directly compared to experimental data.

The   $\gamma p \to \eta p$ and $\gamma n \to \eta n$ 
differential cross sections are compared in Fig.~\ref{eta-diff}. Similar to the photoproduction 
on the proton, the $\gamma n \to \eta n$ reaction is strongly influenced by the $S_{11}(1535)$ 
resonance contributions in the energy region from the threshold and up to 1.6 GeV.  Due to the 
negative sign of the $A^{n}_{1/2}(S_{11}(1535))$ the angular distribution in the latter case has a different 
profile. A similar behaviour has been observed in \cite{Mukhopadhyay:1995}.  At energies above 
1.6 GeV the excitations of  the $S_{11}(1650)$ and $P_{11}(1710)$ states give a strong effect 
which changes the angular distribution at forward angles. Above the 1.750 GeV the contributions from
these states drop rapidly and the  calculated  distribution at 1.835 GeV becomes similar to that of 
$\gamma p \to \eta p$ reaction. 

The calculated $(d\sigma/d\Omega)_n/(d\sigma/d\Omega)_p$ ratio is compared in Fig.~\ref{eta_npr} to
the experimental data from \cite{Hoffmann-Rothe:1997sv}. Unfortunately, because
of the large error bars no solid conclusion about the angular dependence can be drawn from this data. 
In the present calculations we find an almost symmetric angular distribution for this ratio with 
a minimum close to $\theta=80^0$.	

\begin{figure}
  \includegraphics*[width=7cm]{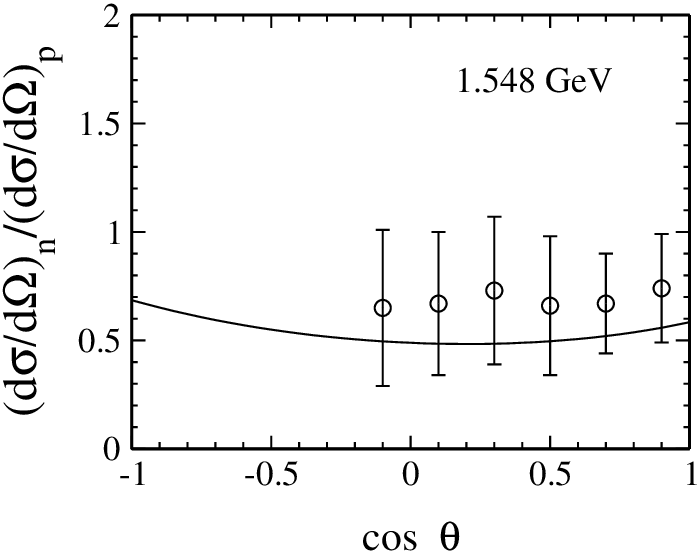}
       \caption{Ratio of the  $\eta$ production cross sections on the proton and the neutron, 
       as a function of the scattering angle and fixed total c.m. energy $\sqrt{s}=$1.548 GeV. 
       The experimental data are taken  from \cite{Hoffmann-Rothe:1997sv}.
      \label{eta_npr}}
  \includegraphics*[width=7cm]{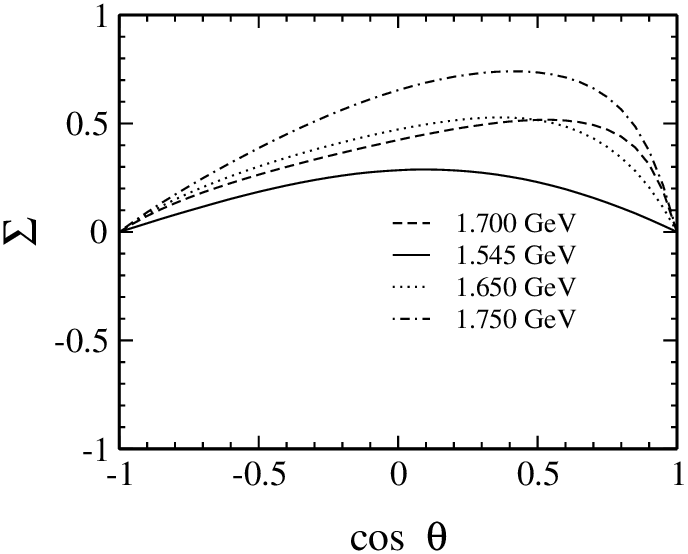}
     \caption{Beam asymmetry as a function of the scattering angle and fixed total c.m. energies.
      \label{eta_asymm}}
\end{figure}

With the parameters fixed as just discussed we now make a prediction for the photon beam asymmetry.
The photon beam asymmetry $\Sigma=\frac{d\sigma_\perp-  d\sigma_\parallel }{d\sigma_\perp+d\sigma_\parallel}$ 
is shown in   Fig.~\ref{eta_asymm} as a function of scattering angle and fixed energies,  
where 
$d\sigma_\perp(d\sigma_\parallel)$ is a differential cross section of the $\gamma n \to \eta n$
reaction with the linearly polarized photons in horizontal (vertical) direction relative to the 
reaction plane.
 The calculated  asymmetry is positive  and has a maximum at forward angles at all energies 
under consideration.

In summary, we have performed  a new coupled-channel analysis of $\eta$ photoproduction in the 
resonance energy region. 
To constrain the  neutron helicity amplitudes of nucleon resonances we include in addition to 
pion photoproduction   the experimental data on the  $\gamma n \to \eta n$ reaction 
from \cite{jaegle:prelim,Hoffmann-Rothe:1997sv}. The inclusion of the $\eta n$ data lead to a significant 
modification of the neutron helicity amplitude of the $S_{11}(1535)$ resonance 
and to a lesser extent - of $S_{11}(1650)$. 
In line with the findings in
\cite{Sauermann:1997,Drechsel:1998,Mukhopadhyay:1995,Chiang:2002vq} we determine the ratio
$A^n_{1/2}(S_{11}(1535))/A^p_{1/2}(N_{1535}^*)=-0.8$ which is required to describe
the $ \eta p$- and  $\eta n$ photoproduction data at energies close to the mass of the 
$S_{11}(1535)$ resonance. Our results show, that the second peak  observed in the $\eta$ photoproduction
on the neutron at 1.66 GeV can be explained by the $S_{11}(1650)$ and $P_{11}(1710)$ resonance
excitations   without invoking  an exotic narrow state.
The differential cross section calculated at the position of the second peak  
shows a rise at backward direction. Above 1.75 GeV the contributions from these resonances
decrease rapidly and the angular distribution becomes similar to that of the $\gamma p \to \eta p$
reaction. 
The precise measurements of the $\eta n$ differential cross section in the energy range 1.6...1.7 GeV 
might be decisive to distinguish between the present coupled-channels calculations 
and results of other theoretical approaches \cite{Azimov:2005jj} and  ~\cite{Fix:2007st}.
 As an interesting by-product of our investigations we predict a beam asymmetry which has been 
measured at GRAAL and in presently being analyzed.

\acknowledgments
We thank I. Strakovsky and L. Tiator for the fruitful discussions and comments.
 The work is supported by Forschungszentrum Juelich.

\bibliographystyle{apsrev}
\bibliography{tau}

\end{document}